\documentclass
[preprint,prb,a4paper,amsfonts,amssymb,floats,titlepage,fleqn,footinbib,lengthcheck,twocolumn,tightenlines,balancelastpage,10pt]{revtex4}%
\usepackage{amsfonts}
\usepackage{amsmath}
\usepackage{amssymb}
\usepackage{graphicx}
\usepackage{longtable}
\usepackage{portland}
\usepackage{supertabular}
\usepackage{array}%
\providecommand{\U}[1]{\protect\rule{.1in}{.1in}}

\begin{document}
\author{Hiroyuki Takeya}
\email{Takeya.Hiroyuki@nims.go.jp}
\affiliation{National Institute for Materials Science,1-2-1 Sengen, Tsukuba, Ibaraki
305-0047, Japan}
\affiliation{JST, Transformative Research-Project on Iron Pnictides (TRIP), Chiyoda, Tokyo
102-0075, Japan}
\author{Mohammed ElMassalami}
\email{massalam@if.ufrj.br}
\affiliation{Instituto de Fisica, Universidade Federal do Rio de Janeiro, Caixa Postal
68528, 21945-970, Rio de Janeiro, Brazil}
\title{Linear magnetoresistivity in the ternary $AM_{2}$B$_{2}$ and $A_{3}$Rh$_{8}%
$B$_{6}$ phases ($A$ = Ca, Sr; $M$ = Rh, Ir)}
\date{\today{}}

\begin{abstract}
We studied the magnetoresistivity of the  $AM_{2}$B$_{2}$ and $A_{3}$Rh$_{8}%
$B$_{6}$  ($A$ = Ca, Sr; $M$ = Rh, Ir) within the ranges 1.8$\leq T\leq$300 K
and 0$\leq H\leq$50 kOe. The zero-field resistivity, $\rho_{\text{0}}(T)$, is
metallic and follows closely the Debye-Gruneisen description. A positive,
non-saturating,\ and dominantly\ linear-in-$H$ magnetoresistivity was observed
in all samples including the ones with a superconducting ground state. Such
$\Delta\rho_{\text{T}}(H)/\rho_{\text{T}}(0)$, reaching 1200\% in favorable
cases, was found to be much stronger for the $AM_{2}$B$_{2}$ compounds and to
decrease with temperature as well as when Ca is replaced by Sr or Rh is
replaced by Ir. Finally, the general features of the observed
magnetoresistivity will be discussed in terms of the Abrikosov model for the
linear magnetoresistivity in inhomogeneous materials.

\end{abstract}
\maketitle

\section{Introduction}

The recent reports of a positive, extraordinarily high, linear
magnetoresistivity (LMR) in nonmagnetic semimetals and semiconductors have
attracted much attentions.\cite{Abrikosov03-RevLinMagRes}$^{-17}$ Extensive
efforts were directed toward the identification of the involved mechanism(s)
as well as toward material optimization for eventual technological
applications\ such as high-density data storage or magnetic sensors and
actuators. LMR was observed over wide ranges of temperatures ($\sim$mK$\leq$
$T\leq$ 400 K) and magnetic fields (few Oe$\leq$ $H\leq$ 600 kOe) and in a
variety of materials such as elemental
metals,\cite{Kapitza29-linearMagRes,Liu98-Bi-FiniteSizeEffects,Yang99-LinMagRes-Bi}
intermetallic
compounds,\cite{Budko98-LMR-RSb2,Young03-LaSb2,Andersson08-Mga2Al3}
Ag$_{2+\delta}X$ ($X$=Se,
Te),\cite{Xu97-LinMagRes-Ag2Te,Husmann02-MR-Ag2Se,Lee02-LMR-Ag2X,vonKreutzbruck09-LMR-Ag2Se}
InSb,\cite{Hu08-Qm-LMR-InSb} Si,\cite{Delmo09-MagRes-Si}
graphene,\cite{Friedman10-LMR-Graphene} graphite,\cite{Morozov05-LMR-Graphite}
GaAs-MnAs,\cite{Johnson10-LMR-MnAs-GaAs} and BaFe$_{2}$As$_{2}$%
.\cite{Tajima10-MR-BaFe2As2}

Classically, a field dependent normalized magnetoresistivity$\ \Delta
\rho_{\text{T}}(H)/\rho_{\text{T}}(0)$ =$\left(  \rho_{\text{T}}%
(H)-\rho_{\text{T}}(0)\right)  /\rho_{\text{T}}(0)$ is quadratic in $H$ for
$\mu_{c}H/c<1$ and saturates for $\mu_{c}H/c>1$ (carrier mobility $\mu
_{c}=e\tau/m^{\ast}$; symbols have their usual meaning). Various
scenarios,\cite{Abrikosov03-RevLinMagRes} with some classical and others
quantum mechanical, were proposed for the interpretation of the deviation of
LMR from the classical prediction.

The so-called Kapitza's LMR is expected in metals, such as Bi, with an open
Fermi surface and a mea-free path which is longer than the\ electronic Larmor
radius.\cite{Kapitza29-linearMagRes,Liu98-Bi-FiniteSizeEffects,Yang99-LinMagRes-Bi}
Another scenario discusses the inhomogeneous conducting media, such as InSb
semiconductor above 200 K: here disorder causes an intermixing\ of the
off-diagonal components of the magnetoresistance MR tensor and, as such, the
associated LMR is due to the distribution in $\mu_{c}$ rather than $\mu_{c}$
itself.\cite{Hu08-Qm-LMR-InSb}

Abrikosov\cite{Abrikosov03-RevLinMagRes}\ identified three classes of
materials wherein quantum LMR can be manifested. First are those homogenous
materials at very low $T$ and strong $H$ and a low concentration of charge
carriers ($n_{c}$) and a small $m^{\ast}$ such that only the lowest Landau
level is populated. Its strength is given by $N_{d}H/(\pi n_{c}^{2}ec)$ where
$N_{d}$ is the concentration of the defect centers. Second are those highly
inhomogeneous materials, e.g. Ag$_{2+\delta}X$ ($X$=Se,
Te),\cite{Xu97-LinMagRes-Ag2Te,Husmann02-MR-Ag2Se,Lee02-LMR-Ag2X,vonKreutzbruck09-LMR-Ag2Se}
wherein metallic inclusions (with higher $n_{c}$) are dispersed within a
matrix having a smaller $n_{c}$, a linear dispersion relation and a vanishing
energy gap. Third are those layered structures such as LaSb$_{2}%
$,\cite{Young03-LaSb2,Budko98-LMR-RSb2} which $-$ due to a particular
configuration of their electronic structure $-$ exhibit a large Fermi surface
(with a classical MR contribution) and, in addition, tiny pockets with a small
effective mass (thus providing a quantum LMR contribution). In this case,
depending on $T$, $H$, and the material properties, LMR may dominate the
magnetoresistive feature.

\begin{figure}[th]%
\centering
\includegraphics[
height=7.7958cm,
width=8.0309cm
]%
{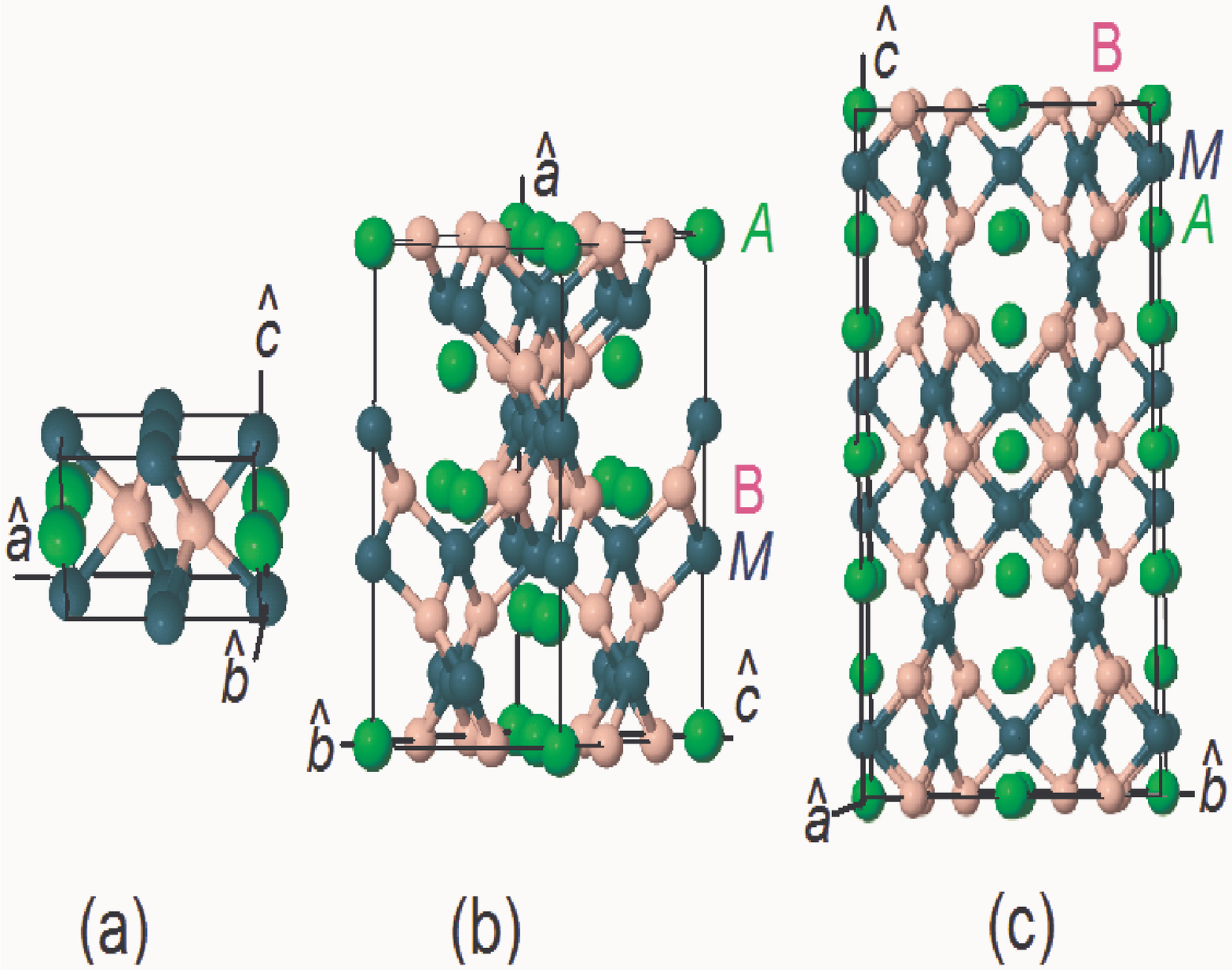}%
\caption{(Color online) The unit-cells, not to scale, of (a) $AM_{3}$B$_{2}$
($P6/mmm$),\cite{ICSD-2009} (b)\ $AM_{2}$B$_{2}$ ($Fdd2$) and (c) $A_{3}M_{8}%
$B$_{6}$ ($Fmmm$) ($A$=Ca, Sr; $M$=Rh, Ir). The structure of $A_{3}M_{8}%
$B$_{6}$ can be visualized as an alternative stacking of the (a) $AM_{3}%
$B$_{2}$ and (b) $AM_{2}$B$_{2}$  sheets.\cite{Jung84-XnRh3n-1B2n} The
stacking direction for $AM_{3}$B$_{2}$, $AM_{2}$B$_{2}$ and $A_{3}M_{8}$%
B$_{6}$ is along the, $c,$ $a,$ and $c$ axis, respectively.}%
\label{structure-12-132-122-386-fig1}%
\end{figure}

In this work, we report on the observation of a relatively strong LMR effect
in the homologous $A_{n}M_{3n-1}$B$_{2n}$ series ($A$=Ca, Sr; $M$=Rh, Ir,
$n$=1, 3). Because the evolution of LMR depends on material parameters such as
$n_{c}$, $\mu_{c}$ and anisotropy, the investigation of MR in different
$A_{n}M_{3n-1}$B$_{2n}$ members (each with its distinct materials properties)
would be helpful in identifying the essential parameters behind the surge of
LMR in this series. In fact, the following three reasons highlight our
interest in studying the functional dependence of LMR in these intermetallics.
First, their structure consists of a combination of alternatively stacked
$AM_{2}$B$_{2}$ and $AM_{3}$B$_{2}$ sheets (see Fig.
\ref{structure-12-132-122-386-fig1}).\cite{Jung84-XnRh3n-1B2n} As such, a
variation in $n$ (e.g. 1 $\leftrightarrow$ 3) entails a variation in the
number of involved layers and, as a consequence, a variation in the electronic
properties. Second, a variation of $A$ (Ca $\leftrightarrow$Sr) or $M$
(Rh$\leftrightarrow$Ir) entails also a possible variation in $n_{c}$, $\mu
_{c}$, or chemical pressure. Third, because both Sr$_{3}$Rh$_{8}$B$_{6}$ and
Ca$_{3}$Rh$_{8}$B$_{6}$ are superconductors while $AM_{2}$B$_{2}$ are
normal\cite{AnM3n-1B2n-Superconductivity} and, furthermore, because all
members show LMR effect, it is interesting to investigate the correlation, if
there is any, between the electronic ground state (whether superconducting or
normal) and their LMR properties.%

\section{Experiment}

Six compounds, $AM_{2}$B$_{2}$ and $A_{3}$Rh$_{8}$B$_{6}$ ($A$=Ca, Sr; $M$=Rh,
Ir), were synthesized using standard solid-state reaction of pure elements in
BN or Ta crucibles (in contrast to the $A_{n}$Rh$_{3n-1}$B$_{2n}$ case, there
is no homologous Ir-based $n$=3 series). Their single phase character was
verified by extensive structural and elemental analyses (see Ref.
19\nocite{AnM3n-1B2n-Superconductivity} for information on the synthesis,
structural and elemental analyses).\ \newline

$\rho(T,H)$ curves of polycrystalline, parallelepiped-shaped samples were
measured by a conventional, home-made, in-line four-point magnetoresistometer.
The geometry of the sample as well as the separation between the contacts were
chosen in such a way as to reduce the contribution of the so-called
geometrical magnetoresistivity, which is a purely geometrical effect
associated with the response to the changing direction of the current carriers
in a magnetic field.\cite{Schroder06-Seminconductors} The longitudinal
geometry ($I\Vert H$) was adopted in most cases so as to avoid such a
geometrical effect and also to detect any possible Shubnikov-de Haas
oscillations.\cite{note-LMR-Absence-ShdeH} As a check, $\rho(T,H)$ of Ca$_{3}%
$Rh$_{8}$B$_{6}$ was measured also along the transversal geometry ($I\bot H$):
as usual,\cite{Hu08-Qm-LMR-InSb,Young03-LaSb2} the transverse $\Delta
\rho_{\text{T}}(H)/\rho_{\text{T}}(0)$ is much higher than the longitudinal
one. The ohmic character was verified within the 1$\leq I\leq$100 mA range for
various samples; measurements reported here were taken with 10 mA. Various
isofield\ and isothermal scans were carried out covering 1.8 $\leq T\leq$300 K
and 0$\leq T\leq$ 50 kOe. The residual resistivity ratio, $RRR=\rho
($300K$)/\rho($1.8K$)$, was found to be $\sim$6 $-$32. $\rho(T,H=0)$ was
considered to be a sum of a residual contribution $\rho_{\text{00}}$ and a
Bloch--Gr\"{u}neisen (BG)
expression,\cite{Allen96-Quantum-Theory-of-Real-Materials}%
\begin{equation}
\rho_{\text{0}}(T)-\rho_{\text{00}}=16\pi^{2}\omega_{\text{D}}\frac{\lambda
}{\omega_{p}^{2}}\left(  \frac{2T}{\theta_{D}}\right)  ^{5}%
{\displaystyle\int\nolimits_{0}^{\frac{\theta_{D}}{2T}}}
\frac{x^{5}}{\sinh(x)^{2}}dx\label{Eq.ResBlochGruneisen}%
\end{equation}
where $\lambda$ is the electron-phonon coupling, $\omega_{p}$ is the Drude
plasma frequency and $\omega_{\text{D}}$ is the Debye phonon frequency. Below,
only $\theta_{D}$ and $\lambda/\omega_{p}^{2}$ are treated as free parameters.

Low-$T$ $\rho_{T}(H)$ isotherms exhibit a strong linear-in-$H$ feature for
$H>H_{cr}$ $\approx$10 kOe ($H_{cr}$ is the crossover field above which the
LMR character dominates): thus for $H>H_{cr}$ and $T<$100 K:
\begin{equation}
\Delta\rho_{\text{T}}(H)/\rho_{\text{T}}(0)\ =a_{\text{0}}+a_{\text{T}%
}.H\label{Eq.RvsH}%
\end{equation}
where $a_{\text{T}}=$ $\left(  \frac{1}{\rho_{\text{0T}}}\frac{\partial\rho
}{\partial H}\right)  _{T}$\ depends on $T$ and the material properties. Plots
of $\Delta\rho_{\text{T}}(H)/\rho_{\text{T}}(0)$ isotherms against $H/\rho
_{o}$ indicate that the Kohler rule is not satisfied. The thermal evolution of
$\Delta\rho_{\text{50kOe}}(T)/\rho_{\text{0}}(T)=(\rho_{\text{50kOe}}%
(T)-\rho_{\text{0}}(T))/\rho_{\text{0}}(T)$ was found to follow the empirical
relation.
\begin{equation}
\Delta\rho_{50\text{kOe}}(T)/\rho_{\text{0}}(T)=b_{\text{H}}.\tanh
(c_{\text{H}}/T)/(d_{\text{H}}.T^{2}+1)\label{Eq.normRt}%
\end{equation}
where the phonon contribution was assumed to be $H$%
-independent,\cite{note-LMR-Phonon-vs-electronic} and $b_{\text{H}}$,
$c_{\text{H}}$ and $d_{\text{H}}$ are sample-dependent parameters\textit{
}that will be used below only for comparative purposes. Above 100 K, this
expression (and its LMR character) was found to be extremely small suggesting
an energy scale of $\sim9$ meV. Finally, for $T\rightarrow$ $\infty$,
$\Delta\rho_{50\text{kOe}}(T)/\rho_{\text{0}}(T)\rightarrow T^{-3}$; on the
other hand, if both Eqs.\ref{Eq.RvsH} and \ref{Eq.normRt} hold as
$T\rightarrow$ 0 K, then $b_{\text{H}}\rightarrow$ $a_{\text{0}}+a_{\text{T}%
}H$: this establishes a link to available theoretical models.

\section{Results}

\subsection{Ca$_{n}$Rh$_{3n\text{-}1}$B$_{2n}$ (\textit{n} =1,3) and
CaIr$_{2}$B$_{2}$}%

\begin{figure}[th]%
\centering
\includegraphics[
height=6.3944cm,
width=7.0973cm
]%
{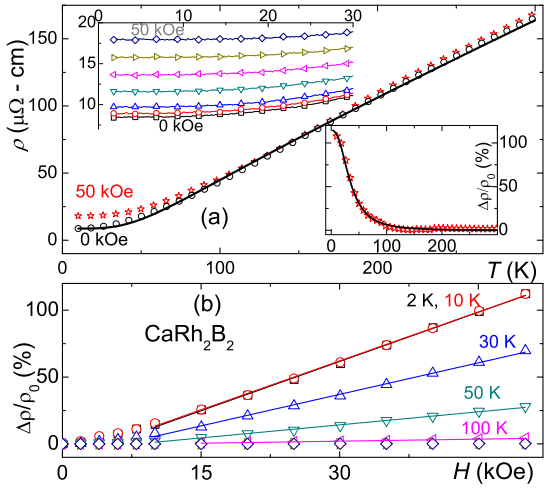}%
\caption{(Color online) $\rho(H,T)$ curves of CaRh$_{2}$B$_{2}$. (a) Isofield
$\rho_{\text{H}}(T)$ curves at $H$=0 and 50 kOe (solid line, $H$=0, is a fit
to Eq. \ref{Eq.ResBlochGruneisen}). \textit{ Inset (Upper-left):} Isofield
$\rho_{\text{H}}(T)$ curves at $H$=0, 5, 10, 20, 30, 40, 50 kOe. \textit{
Inset (Lower-right)}: Thermal evolution of$\ \Delta\rho_{\text{50kOe}}%
(T)/\rho_{0}(T)$ (solid line represents Eq. \ref{Eq.normRt}). (b) $\Delta
\rho_{\text{T}}(H)/\rho_{\text{T}}(0)$ isotherms (solid lines are fits to Eq.
\ref{Eq.RvsH}).}%
\label{RvsT-RvsH-CaRh2B2-Fig2}%
\end{figure}

\begin{figure}[th]%
\centering
\includegraphics[
height=5.4147cm,
width=7.0951cm
]%
{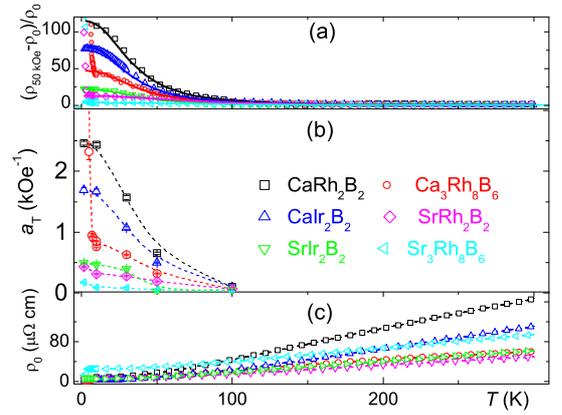}%
\caption{(Color online) Thermal evolution of (a) longitudinal $\Delta
\rho_{\text{50kOe}}(T)/\rho_{\text{0}}(T)$ (solid lines are fits to Eq.
\ref{Eq.normRt}), (b) $a_{\text{T}}$ (based on fit to Eq. \ref{Eq.RvsH}), and
(c) the measured zero-field $\rho_{\text{0}}(T)$. For the three panels, the
anomalous features of $A_{3}$Rh$_{8}$B$_{6}$ ($A$=Ca, Sr) at the lowest
temperatures are related to the onset of
superconductivity.\cite{AnM3n-1B2n-Superconductivity}}%
\label{param-Res-A3M3n-1B3n-Fig3}%
\end{figure}

Figure \ref{RvsT-RvsH-CaRh2B2-Fig2}(a) emphasizes the\ metallic character of
the zero-field $\rho_{\text{0}}(T)$ curve of CaRh$_{2}$B$_{2}$; it follows the
BG description (see Eq. \ref{Eq.ResBlochGruneisen} and Table
\ref{Tab.FitParameters}) emphasizing the dominant strength of the
phonon-electron interaction. In addition, Fig. \ref{RvsT-RvsH-CaRh2B2-Fig2},
in particular the insets, manifests a relatively strong $\Delta\rho/\rho_{o}$,
in which $\Delta\rho_{\text{50kOe}}(T)/\rho_{\text{0}}(T)$ is relatively
strong (%
$>$%
100\%) at low-$T$ but decreases sharply with temperature (below 4\% for
temperatures above 100 K) following approximately Eq. \ref{Eq.normRt} (see
inset of Fig. \ref{RvsT-RvsH-CaRh2B2-Fig2} (a), Fig.
\ref{param-Res-A3M3n-1B3n-Fig3} (a) and Table \ref{Tab.FitParameters}).

\begin{table}[t]
\caption{The parameters $b_{\text{H}}$, $c_{\text{H}}$, and $d_{\text{H}}%
\ $were obtained from fitting longitudinal $\Delta\rho_{\text{50kOe}}%
(T)/\rho_{\text{0}}(T)$ to Eq. \ref{Eq.normRt} [bracketed values represent
transverse geometry]. $\theta$ and $\lambda/\omega^{2}$ are from the BG
fit\ (Eq. \ref{Eq.ResBlochGruneisen}).\ $RRR$ represents $\rho($300K$)/\rho
($1.8K$)$. The high $RRR$ of the $n$=3 members is related to the onset of
partial superconductivity. $\rho_{\text{50kOe}}(T)$ increases with $H$,
$\theta$ (lattice hardening), and $\rho_{\text{0}}$ (residual resistivity).}
\begin{tabular}
[c]{lllll}\hline\hline
$A_{n}$ & parameter & $n=1$ &  & $n=3$\\\hline
&  & Rh & Ir & Rh\\\cline{3-5}%
Ca$_{n}M_{3n-1}$B$_{2n}$ &  &  &  & \\
& $RRR$ & 19.3 & 32.2 & 13.5\\
& $\theta$ $\pm10$K & 280 & 380 & 200[200]\\
& $\lambda\omega_{\text{D}}/\omega_{p}^{2}$ $(\times$10$^{-4}$ $\Omega$-cm$)$
& 19.2 & 18.4 & 5.2[5.1]\\
& $b_{\text{H}}$ & 115 & 80 & 47[85]\\
& $c_{\text{H}}$ & 35 & 30 & 35[33]\\
& $d_{\text{H}}(\times$10$^{-4}$ K$^{-2})$ & 5 & 4 & 3.4[3.4]\\
Sr$_{n}M_{3n-1}$B$_{2n}$ &  &  &  & \\
& $RRR$ & 13.9 & 12.3 & 6.0\\
& $\theta\pm10$K & 300 & 250 & 280\\
& $\lambda\omega_{D}/\omega_{p}^{2}$ $(\times$10$^{-4}$ $\Omega$-cm$)$ & 5.7 &
6.1 & 8.4\\
& $b_{\text{H}}$ & 13 & 22 & 3.5\\
& $c_{_{\text{H}}}$ (K) & 100 & 57 & 90\\
& $d_{\text{H}}(\times$10$^{-4}$ K$^{-2})$ & 1.75 & 4 & 0.7\\\hline\hline
\end{tabular}
\label{Tab.FitParameters}%
\end{table}%

Similar conclusions were drawn from the analysis of various $\rho_{\text{T}%
}(H)$ isotherms, where all $\Delta\rho_{\text{T}}(H)/\rho_{\text{T}}(0)$
isotherms of Fig. \ref{RvsT-RvsH-CaRh2B2-Fig2} (b) manifest a positive MR with
a positive curvature and a predominant high-$H$ LMR character. Fitting
$\Delta\rho_{\text{T}}(H>$10kOe$)/\rho_{\text{T}}(0)$ to Eq. \ref{Eq.RvsH}
gave the parameter plotted in Fig. \ref{param-Res-A3M3n-1B3n-Fig3} (b) which,
once more, emphasizes the strong $T$-dependence of $\Delta\rho_{\text{T}%
}(H)/\rho_{\text{T}}(0)$.
\begin{figure}[th]%
\centering
\includegraphics[
height=6.4317cm,
width=7.0951cm
]%
{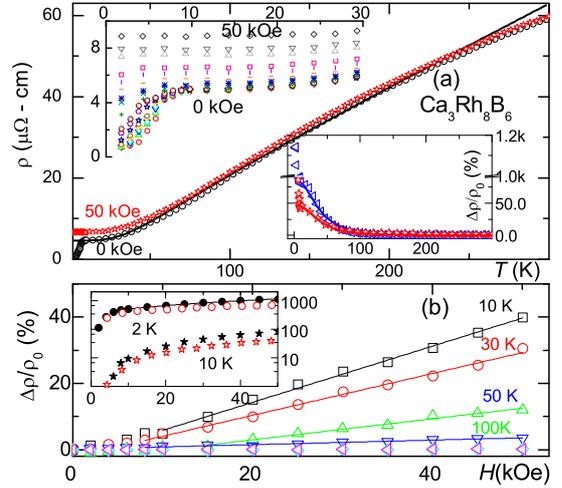}%
\caption{(Color online) $\rho(H,T)$ curves of Ca$_{3}$Rh$_{8}$B$_{6}$. (a)
$\rho_{\text{H}}(T)$ curves at $H$=0 and 50 kOe (solid line, $H$=0, is a fit
to Eq. \ref{Eq.ResBlochGruneisen}). \textit{ Inset (upper-left):} thermal
evolution of isofield $\rho_{\text{H}}(T)$ curves at $H$=0, .., 50 kOe.
\textit{ Inset (lower-right)}: thermal evolution of $\Delta\rho_{\text{50kOe}%
}(T)/\rho_{\text{0}}(T)$ curves (triangles are transversal; stars are
longitudinal; solid lines are fits to Eq. \ref{Eq.normRt}). (b) Longitudinal
$\Delta\rho_{\text{T}}(H)/\rho_{\text{0}}(0)$ isotherms (solid lines are fits
to Eq. \ref{Eq.RvsH}). \textit{Inset: }A semilog plot of the $\Delta
\rho_{\text{T}}(H)/\rho_{\text{T}}(0)$ isotherms at 2 and 10 K: filled (open)
symbols represent transversal (longitudinal) arrangement.}%
\label{RvsT-RvsH-Ca3Rh8B6-Fig4}%
\end{figure}

In contrast to CaRh$_{2}$B$_{2}$, low-$T$ $\rho(H$,$T)$ of Ca$_{3}$Rh$_{8}%
$B$_{6}$ (Fig. \ref{RvsT-RvsH-Ca3Rh8B6-Fig4}) show\ a\ superconducting
state\cite{AnM3n-1B2n-Superconductivity} below $T_{c}\approx$4 K and,
surprisingly, the resistivity within the superconducting phase does not
completely vanish indicating an absence of percolation. Above $T_{c}$, the
normal metallic state follows a BG description (Fig.
\ref{RvsT-RvsH-Ca3Rh8B6-Fig4} (a) and Table \ref{Tab.FitParameters}) however,
for temperatures above 230K, there is a weak deviation, away from Eq.
\ref{Eq.ResBlochGruneisen}. A sizable $\Delta\rho_{\text{T}}(H)/\rho
_{\text{T}}(0)$ is evident in most curves of Fig.
\ref{RvsT-RvsH-Ca3Rh8B6-Fig4}; in particular,\ Fig.
\ref{RvsT-RvsH-Ca3Rh8B6-Fig4}(b) shows that $\rho_{T<T_{c}}(H>H_{c2})$
manifests a negative curvature while $\rho_{T>T_{c}}(H)$ manifests a positive
and almost linear evolution.\ Fitting these curves to Eq. \ref{Eq.RvsH}
yielded $a_{\text{T}}$, the thermal evolution of which is plotted in\ Fig.
\ref{RvsT-RvsH-Ca3Rh8B6-Fig4}.

The normalized $\Delta\rho_{\text{1.8K}}(H)/\rho_{\text{1.8K}}(0)$ reaches, at
50 kOe, an impressive value of 1200\% (see inset of
Fig.\ref{RvsT-RvsH-Ca3Rh8B6-Fig4}(b)): this is attributed to the presence of
the superconducting state (much higher value would be attained if
$\rho_{\text{T}}(0)$ is decreased further).%

\begin{figure}[th]%
\centering
\includegraphics[
height=6.3856cm,
width=7.0973cm
]%
{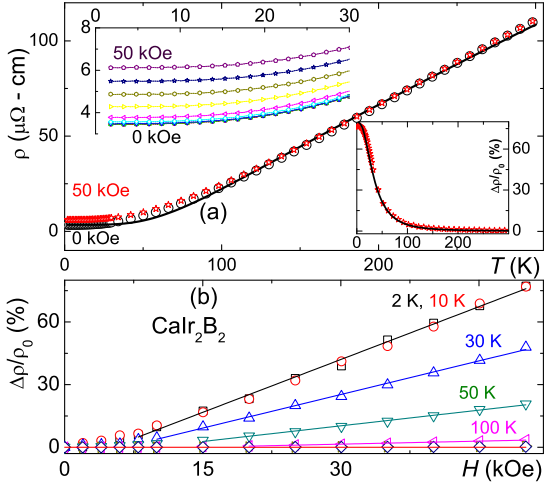}%
\caption{(Color online) $\rho(H,T)$ curves of CaIr$_{2}$B$_{2}$. (a) Isofield
$\rho_{_{H}}(T)$ curves at $H$=0 and 50 kOe (solid line, $H$=0, is a fit to
Eq. \ref{Eq.ResBlochGruneisen}). \textit{ Inset (upper-left):} Thermal
evolution of isofield $\rho_{\text{H}}(T)$ curves at $H$=0, .., 50 kOe.
\textit{Inset }(l\textit{ower-right)}: Thermal evolution of $\Delta
\rho_{\text{50kOe}}(T)/\rho_{\text{0}}(T)$ curve (solid line is a fit to Eq.
\ref{Eq.normRt}). (b) The $\Delta\rho_{\text{T}}(H)/\rho_{\text{T}}(0)$
isotherms (solid lines are fits to Eq. \ref{Eq.RvsH}).}%
\label{RvsT-RvsH-CaIr2B2-Fig5}%
\end{figure}
\ Figure \ref{RvsT-RvsH-CaIr2B2-Fig5} of CaIr$_{2}$B$_{2}$ reflects the same
features that were observed in CaRh$_{2}$B$_{2}$: a metallic $\rho_{\text{T}%
}(0)$ with a BG character (Table \ref{Tab.FitParameters}), a predominant LMR
feature and a relatively strong$\ \Delta\rho_{\text{50kOe}}(T)/\rho_{\text{0}%
}(T)$ effect at low $T$ but decays rapidly at higher $T$, dropping to below
3\% above 100 K. The $a_{\text{T}}$ parameter (the fit of $\rho_{\text{T}}(H)$
to Eq. \ref{Eq.RvsH} $-$Fig. \ref{RvsT-RvsH-CaIr2B2-Fig5}(b)) is shown in Fig.
\ref{param-Res-A3M3n-1B3n-Fig3}.

\subsection{Sr$_{n}$Rh$_{3n-1}$B$_{2n}$(\textit{n}=1,3) and SrIr$_{2}$B$_{2}$}%

\begin{figure}[th]%
\centering
\includegraphics[
height=5.5399cm,
width=6.0934cm
]%
{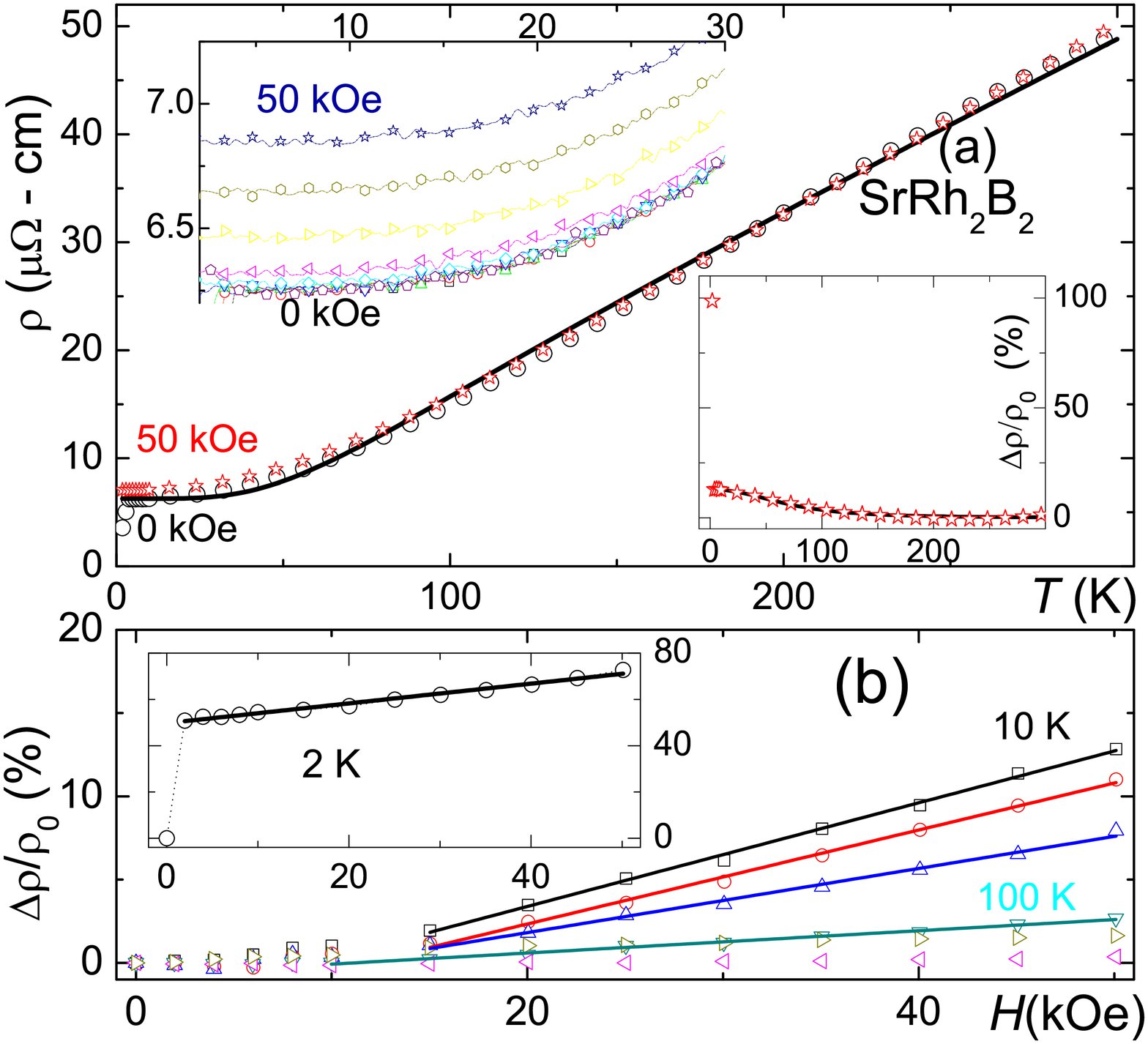}%
\caption{(Color online) $\rho(H,T)$ curves of SrRh$_{2}$B$_{2}$. (a) Isofield
$\rho_{\text{H}}(T)$ curves at $H$=0 and 50 kOe (solid line, $H$=0, is a fit
to Eq. \ref{Eq.ResBlochGruneisen}). \textit{Inset (upper-left):} Thermal
evolution of isofield $\rho_{\text{H}}(T)$ curves at $H$=0, .., 50 kOe.
\textit{Inset (lower-right)}:$\ \Delta\rho_{\text{50kOe}}(T)/\rho_{\text{0}%
}(T)$ curve (solid line is a fit to Eq.\ref{Eq.normRt}). (b) $\Delta
\rho_{\text{T}}(H)/\rho_{\text{T}}(0)\ $isotherms (solid lines are fits to Eq.
\ref{Eq.RvsH}).}%
\label{RvsT-RvsH-SrRh2B2-Fig6}%
\end{figure}

\begin{figure}[th]%
\centering
\includegraphics[
height=6.5064cm,
width=7.0973cm
]%
{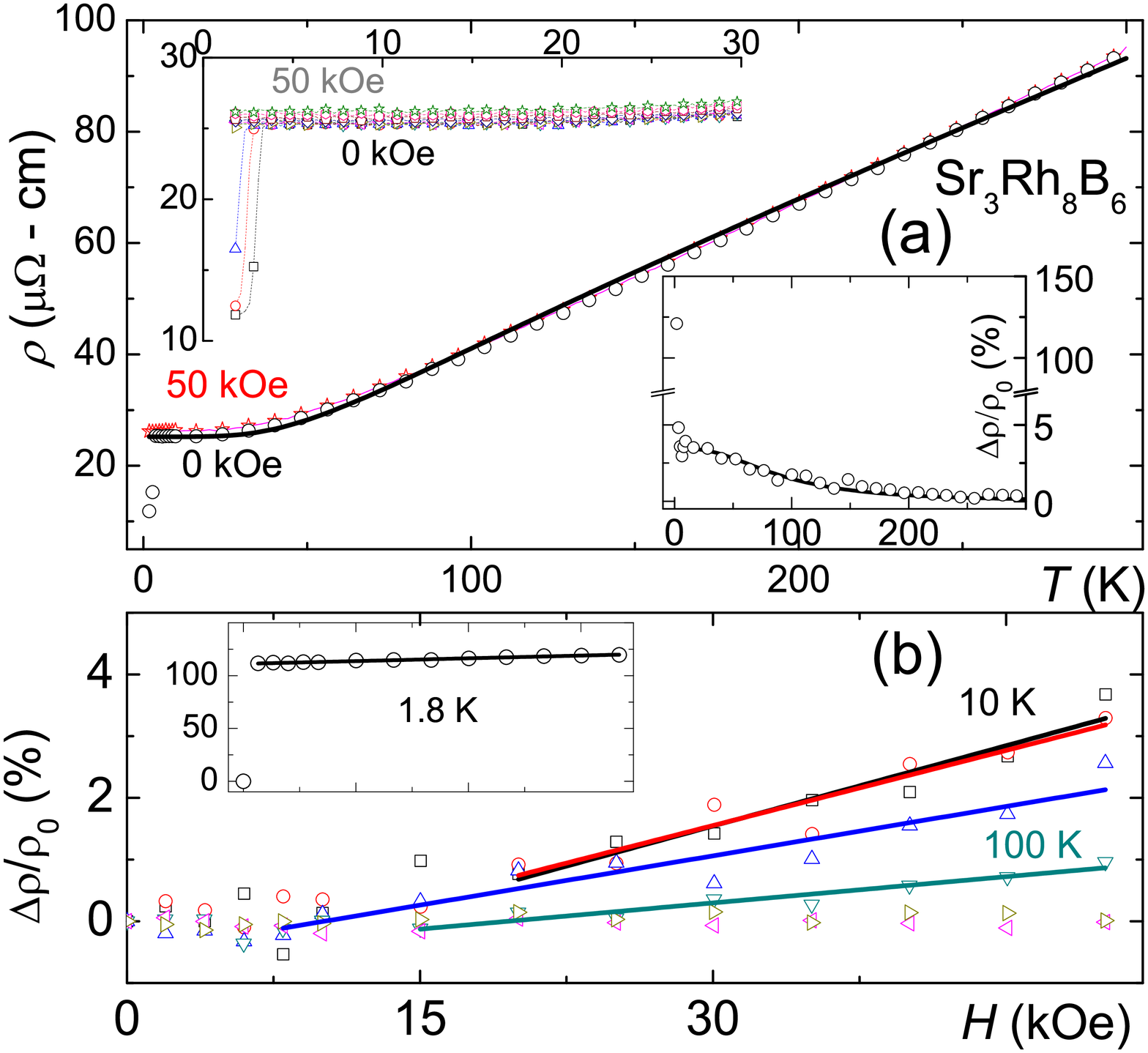}%
\caption{(Color online) $\rho(H,T)$ curves of Sr$_{3}$Rh$_{8}$B$_{6}$. (a)
Isofield $\rho_{\text{H}}(T)$ curves at $H$=0 and 50 kOe (solid line, $H$=0,
is a fit to Eq. \ref{Eq.ResBlochGruneisen}). \textit{Inset (upper-left):}
Thermal evolution of isofield $\rho_{\text{H}}(T)$ curves at $H$=0, .., 50
kOe. \textit{Inset (lower-right )} $\Delta\rho_{\text{50kOe}}(T)/\rho
_{\text{0}}(T)$ curve (solid line is a fit to Eq.\ref{Eq.normRt}). (b)
$\Delta\rho_{\text{T}}(H)/\rho_{\text{T}}(0)\ $isotherms (solid lines are fits
to Eq. \ref{Eq.RvsH}). \textit{Inset: }An expansion of the $\Delta
\rho_{\text{1.8K}}(H)/\rho_{\text{1.8K}}(0)$ isotherm.}%
\label{Rvst-Rvsh-Sr3Rh8B6-Fig7}%
\end{figure}

$\rho(H,T)$ curves of SrRh$_{2}$B$_{2}$ (Fig. \ref{RvsT-RvsH-SrRh2B2-Fig6})
manifest magnetoresistive features that are very similar to those found
in\ CaRh$_{2}$B$_{2}$ except that the strength of the effect is smaller and
there is a weak superconducting secondary phase (namely Sr$_{3}$Rh$_{8}$%
B$_{6}$) which is believed to be behind the drop in the magnetoresistivity
of\ SrRh$_{2}$B$_{2}$ below that of SrIr$_{2}$B$_{2}$ (compare Figs.
\ref{param-Res-A3M3n-1B3n-Fig3} and \ref{RvsT-RvsH-SrRh2B2-Fig6}). On the
other hand, Fig. \ref{Rvst-Rvsh-Sr3Rh8B6-Fig7} shows that Sr$_{3}$Rh$_{8}%
$B$_{6}$ superconducts below $T_{c}\approx$3.5 K, exhibits a BG-type
resistivity above $T_{c}$ and has MR features that are very similar to, but
almost two orders of magnitude weaker than, those of Ca$_{3}$Rh$_{8}$B$_{6}$.%

\begin{figure}[th]%
\centering
\includegraphics[
height=6.5394cm,
width=7.0951cm
]%
{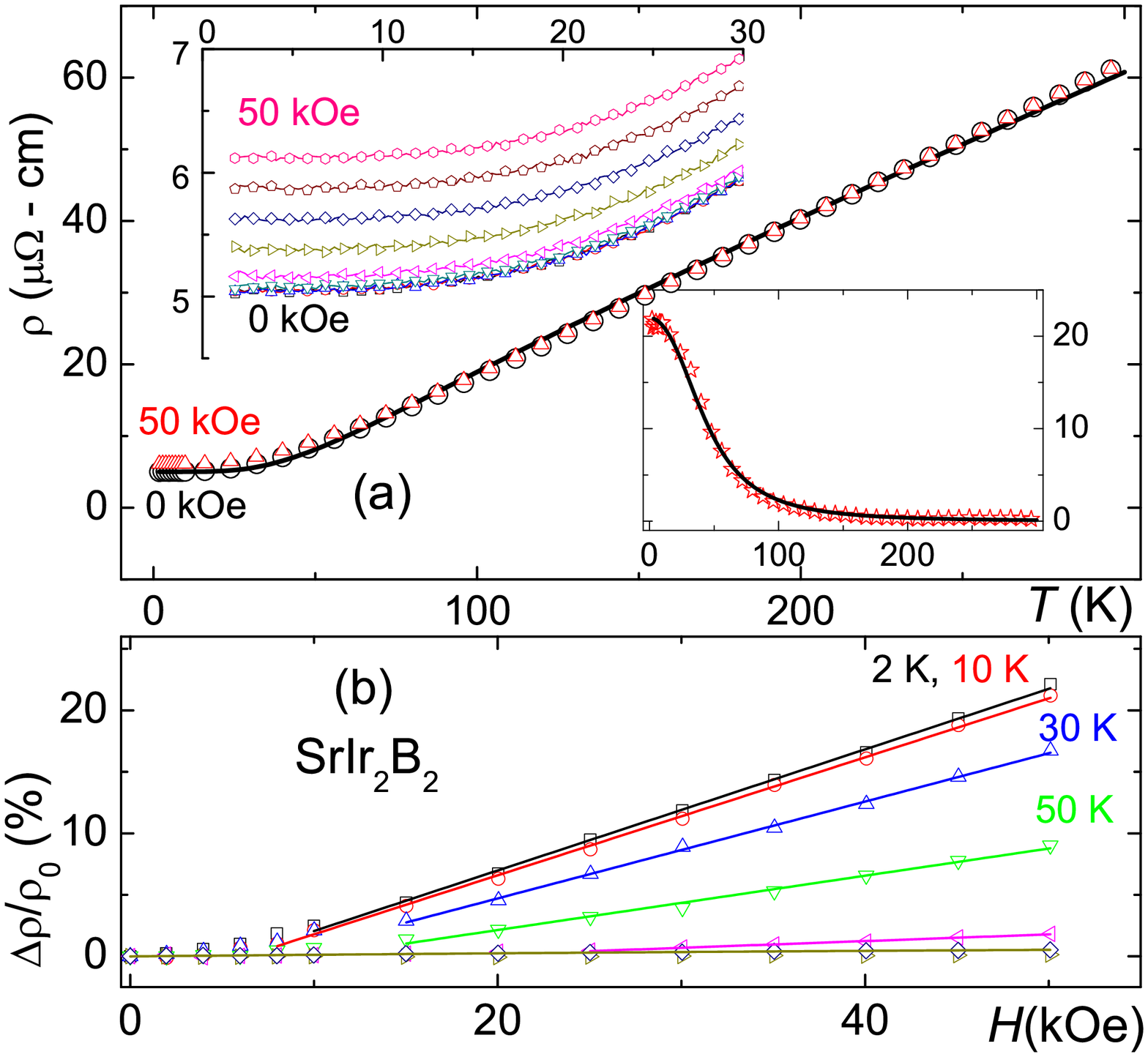}%
\caption{(Color online) $\rho(H,T)$ curves of SrIr$_{2}$B$_{2}$. (a) Isofield
$\rho_{\text{H}}(T)$ curves at $H$=0 and 50 kOe (solid line, $H$=0, is a fit
to Eq. \ref{Eq.ResBlochGruneisen}). \textit{Inset (upper-left ):} Thermal
evolution of isofield $\rho_{\text{H}}(T)$ curves at $H$=0, .., 50 kOe.
\textit{ Inset (lower-right )}: $\Delta\rho_{\text{50kOe}}(T)/\rho_{\text{0}%
}(T)$ curve (solid line is a fit to Eq.\ref{Eq.normRt}). (b) $\Delta
\rho_{\text{T}}(H)/\rho_{\text{T}}(0)$ isotherms (solid lines are fits to Eq.
\ref{Eq.RvsH}).}%
\label{RvsT-RvsH-SrIr2B2-Fig8}%
\end{figure}

Similar to the cases found in Ca$M_{2}$B$_{2}$ isomorphs, $\rho(H$ ,$T)$ of
SrIr$_{2}$B$_{2}$ (Fig. \ref{RvsT-RvsH-SrIr2B2-Fig8}) show all the features
that we mentioned above: the metallic resistivity obeying a BG description,
the predominantly LMR character and the strong $T$-dependence of$\ \Delta
\rho_{\text{50kOe}}(T)/\rho_{\text{0}}(T)$ up to 100 K. From a fit of
$\Delta\rho_{\text{T}}(H)/\rho_{\text{T}}(0)$ to Eq. \ref{Eq.RvsH}, we
obtained $a_{\text{T}}$ (given in Fig. \ref{param-Res-A3M3n-1B3n-Fig3}).%

\section{Discussion and Conclusions}

Our experiments indicated that $\Delta\rho_{\text{T}}(H)/\rho_{\text{T}}(0)$
of $A_{n}M_{3n-1}$B$_{2n}$ is positive, non-saturating and dominantly linear
above $\sim$10 kOe, and that a relatively strong $\Delta\rho_{\text{T}%
}(H)/\rho_{\text{T}}(0)$ was observed in both superconducting and normal
members, which decreases sharply with temperature and whenever $n$ is
increased, Rh\ is replaced by Ir, or Ca is replaced by Sr. These features as
well as other magnetoresistive properties of the studied members were compared
in Fig. \ref{param-Res-A3M3n-1B3n-Fig3} and Table \ref{Tab.FitParameters};
evidently the Ca-based isomorphs exhibit a higher $RRR$, a higher
$\lambda/\omega_{p}^{2}$ and a higher $\Delta\rho_{\text{T}}(H)/\rho
_{\text{T}}(0)$.

It is recalled that a reduction in the LMR is usually related to an increase
in $n_{c}$, $\mu_{c},$ or a decrease in $N_{d}$ (for a quantum LMR, a smearing
of the Landau levels).\cite{Hu08-Qm-LMR-InSb} In turn, the variation in any of
$n_{c}$, $\mu_{c}$, or $N_{d}$ ($\rho_{0}$) can be straight forwardly
associated with a related variation in $T$, $H$ or the material properties.
Along this line of arguments, we discuss the above-mentioned MR features of
$A_{n}M_{3n-1}$B$_{2n}$ series.

First, the drop in $\Delta\rho/\rho_{\text{0}}$ with increasing $n$ is
attributed to an increase in $n_{c}$. Because the structure of $A_{3}M_{8}%
$B$_{6}$, in contrast to $AM_{2}$B$_{2}$, includes additional, sandwiched
$AM_{3}$B$_{2}$ layers (Fig. \ref{structure-12-132-122-386-fig1}), it is
inferred that the introduction of $AM_{3}$B$_{2}$ enhances $n_{c}%
$.\cite{note-LMR-AM3B2} In fact, rewriting the $n$=3 member as $A_{1}M_{8/3}%
$B$_{2}$ already suggests that this enhancement is due to a contribution from
the $4d^{8}5s^{1}$-subbands of the extra Rh. Such a higher $n_{c}$ is
consistent with the surge of superconductivity in the $n$=3
members.\cite{AnM3n-1B2n-Superconductivity} Second, the fact that the
resistivity within the superconducting state of $A_{3}$Rh$_{8}$B$_{6}$ does
not vanish is an indication that these $n$=3 samples contain superconducting
regions dispersed within a nonsuperconducting matrix. This feature excludes
the applicability of the classical LMR\ models; rather it supports the
Abrikosov LMR scenario for inhomogeneous media.\cite{Abrikosov03-RevLinMagRes}

By generalizing this inhomogeneous configuration to the $n$=1
members\cite{note-LMR-inhomogeneity} and assuming the variation in the LMR
effect to be related to a corresponding variation in either $n_{c}$ or carrier
dynamics (influenced by pressure, charge doping, $T$, or $H$), the
above-mentioned experimental results can be satisfactorily explained. As an
example, the fact that $\Delta\rho/\rho_{\text{0}}$ of Sr-based compounds are
lower than their Ca-based isomorphs is attributed to a negative chemical
pressure which is induced by the substitution of isovalent, relatively
large-sized Sr$^{+2}$ into the Ca$^{+2}$ site. Similarly, the reduction of
$\Delta\rho/\rho_{\text{0}}$\ caused by the replacement of Rh by Ir
($5d^{7}5s^{2}$) is attributed to an increase in $n_{c}$ that overwhelms the
influence of an increased antisymmetric spin-orbit interaction. It is recalled
that the space group of $AM_{2}$B$_{2}$ is $Fdd2$ (having no inversion
symmetry operator)\cite{AnM3n-1B2n-Superconductivity} while the space group of
$A_{3}M_{8}$B$_{6}$ is $Fmmm$ (with an inversion symmetry operator).
Accordingly, the antisymmetric spin-orbit interaction in the former series
would exercise a considerable influence (via a spin splitting of the
quasi-particle states) on the electronic
properties.\cite{Mineev09-Sup-NonInversion} According to
Abrikosov\cite{Abrikosov03-RevLinMagRes}, a linear spectrum may arise due to
an absence of a symmetry inversion centre. Because a linear spectrum implies a
smaller effective mass, the absence of a symmetry inversion would enhance the
quantum LMR of the $n$=1 members. Finally, the thermal rate of decrease of
$\Delta\rho_{\text{T}}(H)/\rho_{\text{T}}(0)$ in\ $A_{n}M_{3n-1}$B$_{2n}$ is
much faster than that of, say, Ag$_{2+\delta}X$ ($X$=Se, Te) (Ref.
8\nocite{Xu97-LinMagRes-Ag2Te}) but similar to that of LaSb$_{2}$ (Ref.
6\cite{Young03-LaSb2}): as $n_{c}$ hardly varies below 300 K, this thermal
decrease is attributed to the phonon-driven decrease in $\mu_{c}$ and a
smearing of the Landau levels.\cite{Johnson10-LMR-MnAs-GaAs}

In summary, a positive, nonsaturating and dominantly linear MR was observed in
the $A_{n}M_{3n-1}$B$_{2n}$ series ($A$=Ca, Sr; $M$=Rh, Ir, $n$=1, 3). This
effect was found to decrease whenever $n$ is increased, Ca is replaced by Sr,
Rh is replaced by Ir, or the temperature is raised. Comparative MR studies
among the different members suggest that LMR can be described by the Abrikosov
model for inhomogeneous media.

\begin{acknowledgments}
We acknowledge partial financial support from the Japan Society for the
Promotion of Science and the Brazilian agencies CNPq and Faperj.
\end{acknowledgments}

\bibliographystyle{apsrev}
\bibliography{add-in-1,ASOC-Unconv-Sup,borocarbides,crystalography,intermetallic,magres,massalami,notes,To-Be-Published}

\begin{thebibliography}{27}
\bibitem[{not({\natexlab{a}})}]{note-LMR-Absence-ShdeH}
\bibinfo{note}{Due to the limited temperature range and disorder in the
  samples, we were unsuccessful in observing Shubnikov-de Haas oscillations.}
\bibitem[{not({\natexlab{b}})}]{note-LMR-Phonon-vs-electronic}
\bibinfo{note}{LMR is stronger at temperatures where phonon density is
  extremely small.}
\bibitem[{not({\natexlab{c}})}]{note-LMR-AM3B2}
\bibinfo{note}{The magnetoresistivity of $AM_{3}$B$_{2}$ is expected to be very
  small.}
\bibitem[{not({\natexlab{d}})}]{note-LMR-inhomogeneity}
\bibinfo{note}{$A_{n}$Rh$_{3n-1}$B$_{2n}$ ($ A$=Ca, Sr), being incongruent
  phases, were precipitated, via peritectic reactions, from an $A$ rich
  composition. Therefore some excess $A$ melts may form around the desired
  phase or being segregated elsewhere.}
\end{thebibliography}

\end{document}